# DeepFork: Supervised Prediction of Info Diffusion

**Ramya Akula[1], Niloofar Yousefi[2] and Ivan Garibay[3]**
Industrial Engineering and Management Systems
University of Central Florida
Orlando, USA
[Ramya Akula](), [Niloofar Yousefi](), [Ivan Garibay]()

## Abstract

Information spreads on complex social networks extremely fast, in other words, a piece of information can go viral within no time. Often it is hard to barricade this diffusion prior to the significant occurrence of chaos, be it a social media or an online coding platform. GitHub is one such trending online focal point for any business to reach their potential contributors and customers, simultaneously. By exploiting such software development paradigm, millions of free software emerged lately in diverse communities. To understand human influence, information spread and evolution of transmitted information among assorted users in GitHub, we developed a deep neural network model: "DeepFork", a supervised machine learning based approach that aims to predict information diffusion in complex social networks; considering node as well as topological features. In our empirical studies, we observed that information diffusion can be detected by link prediction using supervised learning. DeepFork outperforms other machine learning models as it better learns the discriminative patterns from the input features. DeepFork aids in understanding information spread and evolution through a bipartite network of users and repositories i.e., information flow from a user to repository to user.

## 1. Introduction

Traffic on social networking sites had drastically increased in the recent past, GitHub is akin. GitHub is a popular online code hosting service that provides an open development environment and visibility of project activity to subscribers by maintaining transparency and collaboration in software development activities (Kalliamvakou et al. 2015), (Dabbish et al. 2012). Crowd-Based Software Engineering (CBSE) allows anyone with domain knowledge to participate in software development tasks:designing, programming, testing, deploying and documenting, widely of late (Cosentino et al. 2017). Before we delve into the topic, we portray the terminology used in GitHub as follows: (i) **Repository**: A folder space to place all project related files, along with the revision history backlogs for the files. These are containers of information(*code*). (ii) **User:** An owner of the repository; a valid contributor who has commit permission to the main repository. (iii) **Fork:** a copy of a repository;this event allows users to experiment with their copy, without affecting the original repository. (iii) **Watch:** Watch enables users to get notified about every change that happens to a repository such as commits, issues, and reports. Information diffusion does happen among assorted users and repositories in GitHub, and this pervasive phenomenon is of interest to those who study complex social networks, although the concept of diffusion of information is borrowed from other fields:epidemiology and economics. Despite of continuous on-going research on workflow analysis of GitHub, there is no concrete model to depict how and why human influence on information propagation happen. Defense Advanced Research Projects Agency (DARPA) has solicited innovative research in the area of computational simulation of the spread and evolution of information in the online environment, this project a.k.a "Computational Simulation of Online Social Behavior (SocialSim)". In the developmental phase of an innovative technology for high-fidelity computational simulation, we first should understand the difference between online communication on traditional social media platforms to social coding platform. We narrowed down our research path to information flow in network structure by the following questions:
*RQ 1: How does information spread over time in complex social coding environment?*
*RQ 2: Given a bipartite network of users and repositories, how can the diffused information evolve?*
*RQ 3: What is the influence of the actions performed by some users on their peers?*

We attempt to answer these questions, by posing the scenario as a link(*information diffusion*) prediction problem on network structure, which is casted as a binary classification task. GitHub data is used in this work, where we consider users and repositories, along with the *followee-follower* relationship among users, forks and watch events between users and repositories to build our network structure. We developed a deep feedforward neural network model: "*DeepFork*"; a supervised machine learning based approach. Our DeepFork model is a direct contribution to SocialSim project and the code is available online[1]. DeepFork is trained with both node and topological features of network to predict information diffusion. To evaluate the performance of DeepFork, we compare it against the benchmark machine learning classifiers and DeepFork performed fairly well. Moreover, we perform an ablation study by ignoring the watch events, to inspect if user awareness in "Watch"-ing the repository helps in predicting whether or not that user will fork that repository.

In the next section, we discuss the related works in areas of link prediction, information diffusion and supervised machine learning and also analysis on Fork events in GitHub. Section 3 explains, the proposed approach along with problem formulation, datasets and data preparation, feature extraction and the model description. Section 4 presents, the experimental setup for empirical study using performance metrics. Section 5 details the results and analysis of the developed model, followed by the conclusion and future scope for extension in section 6.

## 2. Related Works
### 2.1 GitHub
Although GitHub is originally launched for software development, through millions of repositories that are freely available, it has also become a knowledge sharing resource around the world for both industry and academia. Forking is usually the first step towards contributing to a project, by modifying a copy of the original project. By the fork event, developers will have a privilege to access the code and use it as and when needed. Based on the study conducted by (Jiang et al. 2013), developers perform the fork event mainly to contribute to the original repository primarily based on their programming language preference. In the later studies done by (Jiang et al. 2017), to address why and how developers fork, (Zhang et al. 2017) showed that the recommendation of repositories based on user preference has made a significant difference in the overall software development trends, while the open source software profited the most. These recent studies motivated us to learn more about how a piece of information is diffused among assorted repositories and users in GitHub.

### 2.2 Information Diffusion Prediction
We adopt definition of diffusion of information from (Rogers et al. 2010), i.e., the transfer of information from one entity to another that can happen through certain channels over a period of time, irrespective of that information being novel or not. There are many existing models in literature for information diffusion, but for this work, we focus on machine learning based link prediction models as well the models that use network structure for link prediction.

### 2.2.1 Network Structure-based Link Prediction:
From past few decades, another emerging line of study in complex social networks is link prediction, ranging from node pair similarities to learning the features of the nodes itself. This kind of study is grounded on a simple logic that any two nodes with compelling similar properties tend to have an interaction between them. When we think of user similarity and social networks, the first thought that occurs to the mind is the user profile. (Bhattacharyya et al. 2011) and (Anderson et al. 2012), conducted an experimental study to acquire knowledge on user similarity by analyzing user profiles and their corresponding interests groups. The actions performed such as editing the article, answering the questions by those users in Stack Overflow and Wikipedia, are recorded into a vector and then cosine similarity between the vectors are calculated to get the node pair match. In extension to this, (Akcora et al. 2013) regulated an exploratory analysis and believed that the actions performed by the users and their corresponding attributes can reflect the personal interest and social behaviors, and hence there is a likelihood of new connections to form among those users. On the other side, the study on topology also gained focus after (Nowell and Kleinberg 2007) graph structure-based work by analyzing the "proximity" of nodes in a network for link prediction. Metrics used for link prediction that is purely based on the topological information are classified into neighbor based, path based and random walk based metrics by (Wang et al. 2015). Readers can refer to this survey, since describing these measures and metrics is not the main focus of the paper. However, we did use a few measures from that list in our experiments. On the sidelines in understanding human behavior on social networks as to why they get connected, theory (Qui et al. 2010) captured growth dynamics based on behavior evolution awareness, and this knowledge of dynamics was later used by (Qui et al. 2011) for link prediction. Furthermore, parallelly, (Li et al. 2011) measured centrality of the nodes for link prediction based on the maximal entropy on random walks, while (Yang et al. 2011) theory was based on homophily nature in social networks and then followed by (Dong et al. 2012) which studied social patterns based on degree distribution. In addition to network structure based and social theory based link prediction approaches, machine learning based approach is another kind. In machine learning based approach, hand crafted features are pick based on the problem, and these feature information is used for learning. We follow this machine learning based approach in this work, thus in the next section, this genre of models are discussed.

---
[1] https://github.com/akula01/DeepFork

## 2.2.2 Machine Learning based Link Prediction:

In the initial exploratory studies performed for machine learning-based link prediction task by (Pavlov et al. 2007) and (Wohlfarth et al. 2008), co-authorship network is used and an advent of semantic features such as research topics, abstracts of the papers and the event information was used in co-authorship graph for binary classification. During the same time, (Scripps et al. 2008) came up with an alignment for the adjacency matrix of a network, based on the topological and node attribute features. (Kunegis et al. 2009) proposed graph transformation; a learning framework for link and edge weight prediction, by reducing high dimensional problem to one-dimensional curve fitting problem. Later, (Cao et al. 2010) proposed a transfer learning idea, i.e., information is transferred through heterogeneous tasks adaptively without neglecting the similarities among these tasks. Here the prediction of links between the users and also the type of items are accredited as a collective link prediction problem. In another study, (Lu et al. 2010) proposed a supervised learning framework that can efficiently and effectively learn the dynamics of the social networks with path-based features for link prediction in an auxiliary network setup. Social imbalances and attitudes that are reflected through weighted links are proposed (Chiang et al. 2011), which was later extended (Leskovec et al. 2010) in conjunction with long cycles in the network supervised learning approach for link prediction. (De et al. 2011) investigated that weighted network that expressed the strength of interaction between the nodes, improved link prediction results in co-authorship network. Another variation in supervised learning framework is (Scellato et al. 2011), which explored features, not only based on common neighbors, but also on global and place features, i.e., the geographic distance between the nodes and also check-in information of common place, respectively. Recently machine learning based link prediction is used in recommender systems, for example (i) User-based features such as education qualification, titles of the books, age of authors and keywords used by those users, aided for predicting the link between user and the product purchased;with the help of supervised learning in a bipartite graph kernel based environment(Li et al. 2013). (ii) For recommending patent partners into a factor graph model which is an interactive supervised learning framework that transits from potential solicitant selection to solicitant refinement based on ranking, is developed; an interactive learning updates the model depending on feedback received(Wu et al. 2013). Another stream that caught the attention of graph theorists and network scientists is predicting information diffusion, which is a primary focus for us as well. (Bourigault et al. 2014) presented social network embedding algorithm to map the observed information diffusion process into a heat diffusion process modeled by a diffusion kernel in the continuous space, where the latent space clearly explains cascades in the training set. Recently, few sophisticated supervised learning architecture are modeled: (i) DeepCAS: at a given time interval, prediction of increase in the cascade size with the help of cascade features and network structure. The features are passed to gated recurrent unit(Chung et al. 2017), where it learns path embeddings while helping in identifying cascade size increment in an end to end deep learning model(Li et al. 2017); (ii) DeepHawkes: is another end to end deep learning-based framework that attempted to fill the gap between prediction and understanding of information cascades. The three components for the Hawkes process are user influence, self-exciting mechanism, time decay effect (Cao et al. 2017); (iii) DeepInf: for social influence prediction, where both user-specific features and network structures incorporated into a convolutional neural and attention networks (Qui et al. 2018).

Background knowledge of all these models and frameworks aid us in modeling of DeepFork, yet answering our research questions. Our problem is showcased as a prediction of information diffusion through binary classification in GitHub network, which is a variant of a bipartite network consisting of two groups, i.e., users and repositories, where there are also edges between users, and information diffuses from user to repository to user to that predominance. To the best of our knowledge, there are no existing link prediction methods, which can directly be applied to such a network structure. Nevertheless, DeepFork is bridging the gap in fusion literature of network structure and information diffusion.

## 3. DeepFork

The aim of this work is to predict information diffusion in complex social coding networks:GitHub. In this section, we describe the problem followed by the specifications of the dataset along with its preparation. Next, detailed description on feature extraction and model development is provided.

### 3.1 Problem Formulation

Consider a directed graph, $G = (U, V, E_{UU}, E_{UV})$, where $U, V$ are the set of nodes representing users and repositories, respectively. Also, let $E_{UU}$ and $E_{UV}$ denote the set of edges connecting a user with another user and a user with a repository. The problem we attempt to solve is to predict a new link between a user and a repository $E^*_{UV}$, which would lead to the diffusion of information. An edge between a user and a repository represents a Fork Event i.e. when a repository is forked by the user, and an edge between two users represent the *follower-followee* relationship between them. A peculiarity of this graph is that there are no edges between repositories themselves. We define information diffusion as an event when a repository is forked by a user at time $T$, is also forked by a *follower* at time $T+1$. This results in a triangular network structure consisting of three nodes: *followee, follower* and the *repository* as shown in Figure 1. The new link we aim to predict represents diffusion of information (*code*) from one user to another via *follower - followee* relationship. We pose the problem of information diffusion prediction as a binary classification problem, and attempt to solve it by a supervised learning approach. In this work, we employ a neural network model to predict information diffusion. This model is trained with features of all the actors involved: *follower*, *followee* and *repository*. To be more specific, we predict the presence or absence of a link between a *follower* and a *repository*, is based on *follower* profile, *followee* profile who "Fork"ed a repository and properties of that "Fork"ed repository. The probability of the *followee* forking a repository based on the popularity of that repository is not explicitly modeled in our approach, nevertheless we expect the model to learn that from the *repository* profile; the expectation is same for watchevents, as shown in Figure 1.

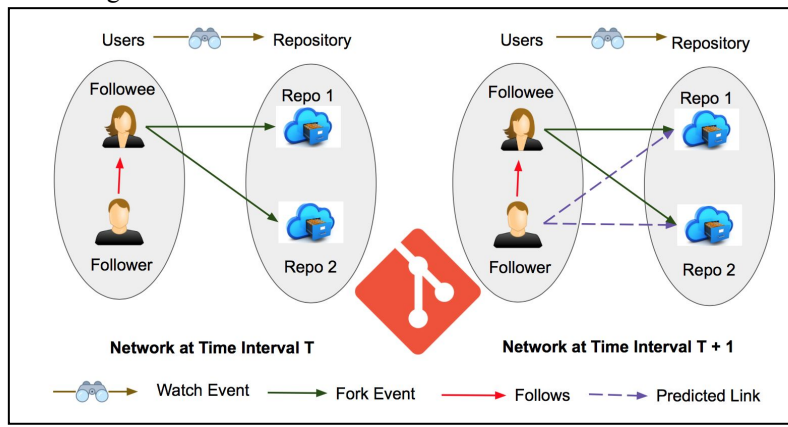

Figure 1. Watch and Fork Events in GitHub; Prediction of link between the *follower* and the *repository* forked by the *followee*, indicating information diffusion(*Code in Repository*) from *followee* to *follower*.

### 3.2 Data Preparation

DARPA furnished an anonymized "GitHub" dataset for Computational Simulation of Online Social Behavior(SocialSim) from GHTorrent for our study. The GHTorrent is a database that backlogs GitHub public event timeline every day and is made available to public for open exploration. Using elasticsearch database from Amazon Web Services and a representational state transfer interface(Battle et al. 2008), we treated "Fork

Event" logs from August 2016 to February 2017. The log entry for each fork event contains an unique ID of the forked repository, and the user, who forked it, followed by the timestamp of the event. The profiles of the users contain information such as his unique ID, number of his followers, number of users he is following, number of repositories he created and so on as shown in Figure 2. The profiles of the repositories contain information such as name, textual description, date of creation, number of forks, number of issues, number of watchers and so on as shown in Figure 3. Along with all this information the database also provides *follower-followee* relation data.

```
{
    'public_repos': 3,
    'extension': {
        'city': None,
        'deleted': False,
        'lon': 0.0,
        'created_dow': 'Thursday',
        'state': None,
        'country_code': None,
        'fake': False,
        'lat': 0.0
    },
    'login_h': 'wCAvKqyEnNvZYL8MrClSJw',
    'created_at': '2016-07-07T02:01:38Z',
    'site_admin': False,
    'ght_id_h': 'jZHv7_5neXJMLD65fjGWGA',
    'followers': 0,
    'location': None,
    'following': 0,
    'type': 'User',
    'company': None
}
```

```
{
    'fork': False,
    'full_name_h': 'wXbHeaRxfwlUjU6tLSqO6g/sau6I1xwK8Gwyl5SHbiTFA',
    'language': None,
    'created_at': '2015-03-18T18:05:11Z',
    'forks_count': 1,
    'updated_at': '2016-02-15T15:06:10Z',
    'ght_id_h': 'kqfkOrFRKEberBIlv8THmw',
    'extension': {
        'created_dow': 'Wednesday',
        'deleted': False,
        'updated_dow': 'Monday'
    },
    'watchers_count': 0,
    'owner': {
        'ght_id_h': 'bhwF7RAhstCbOr5hENzTug',
        'login_h': 'wXbHeaRxfwlUjU6tLSqO6g',
        'type': 'User'
    },
    'name_h': 'sau6I1xwK8Gwyl5SHbiTFA',
    'issue_count': {
        'open_issues_count': 1,
        'total_issues_count': 1
    },
    'description_m': 'aplicativo para etec'
}
```

Figure 2(left) Figure 3(right) are sample user profile and repository profile respectively.

To extract the data required to train our neural network model, we need to pre-process the Github event logs, where each log entry contains details such as the user, who forked a repository; and repository, which got forked. We split the data into two-time intervals $T_1$ (Aug 1, 2016 - Dec 31, 2016) and $T_2$ (Jan 1, 2017 - Feb 1, 2017), and the data from $T_1$ & $T_2$ are used to train and test our model, respectively. For our training purpose, we consider all the repositories that were forked at least 25 times, along with all users, who have forked at least 10 repositories within $T_1$. The reason behind this data filtration is to consider only those users and repositories that are active and contribute to information diffusion. The resultant network has 74,393 nodes of which 21,474 nodes are repositories and 52,919 nodes are users, along with 7,69,928 edges connecting these nodes. From this network, we extract triplets of (*followee, follower, repository*). These triplets are considered positive if we observe information diffusion, i.e., if the repository is forked by both the *follower* and the *followee* and the rest are considered the negative sample as shown in Figure 4. 20% of the training data is used for validation.

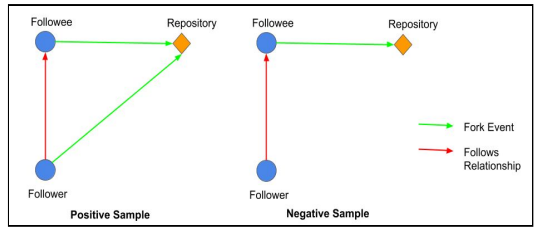

Figure 4. Positive and Negative Samples

### 3.3 Feature Extraction

Once the positive and negative samples are collected, feature vectors are built from these samples to train the model. Each data sample is a triplet with two user IDs and one repository ID. From the user ID, we get the user profile and extract the following features: number of followers, number of followees, number of repositories created by the user, user creation timestamp and if the user is an administrator. Similarly, from the repository profile, the features extracted are: number of fork events, number of issues, number of open issues, number of watchers and repository creation time stamp. As there are five features extracted from both the user and repository profiles, the size of the full feature vector for a data sample is **15**. Along with these 15 data points we also consider adding two binary values indicating if the *followee* and the *follower* are watching the repository, making the final size of the feature vector for training as **17**. Each sample is also associated with a binary label indicating if the sample triplet is positive or negative. For experiment, these 17 node features in the first set and 6 topological features using node similarity measures are used for model training. We compute the 6 node similarity measures for the edges (*followee, repository*) and (*followee, follower*), and use this resultant **12** dimensional feature vector to predict the link between (*follower, repository*). Below is the brief description of the chosen similarity measures.
- **Katz:** This measure takes all the paths between two nodes into the count, where bigger weights are allocated to shorter paths over the longer(Katz and Leo 1953).
- **Common Neighbor**: As the name indicates, given any two nodes the common neighbor is the one that has direct interaction with both these nodes (Newman and Mark 2001).
- **Preferential Attachment**: Any node that has a higher number of prior incoming edges will most likely get connected by new nodes; Rich get richer phenomenon (Barabasi et al. 2002).
- **Adamic Adar:** This measure refines the simple counting of common features by weighting rarer features more heavily (Adamic et al. 2003).
- **Jaccard Coefficient:** This measures is a commonly used when two nodes **X** and **Y** have a neighbor **N**, for a randomly selected neighbor **N** that either **X** or **Y** has(Jaccard and Paul 1901).
- **Total Neighbors**: This measure counts all the neighbors for any given node.

These measures are originally aimed to find similarity between nodes in variety of scenarios, however lately adopted for social network analysis.

### 3.4 Model Description

For learning to predict the links resulting in information diffusion Deep Neural Networks (DNN) are used. The input to our model is a d-dimensional feature vector $X_i = [x_1, x_2,...., x_d]$, as explained in Section 3.3. In a DNN, the input is passed through a series of layers where at each layer, the input $X_i$ is multiplied by a weight matrix W, to produce an output $X_o$. The elements of the weight matrix $W_{jk}$ are the individual weights between nodes in the previous layer(**j**) and the next layer(**k**). The outputs from a layer are then passed through a non-linear activation function such as sigmoid or rectified linear units before moving on to the next layer. At the final layer, the output of the model is compared with the expected output and the loss is calculated. The weights in the model are updated by a factor a.k.a learning rate, using gradient descent back-propagation algorithm to minimize the loss of all training samples.

### 3.5 Benchmark Methods

In this section, we describe the standard machine-learning models that we elected to compare the performance of our DeepFork model. To that end, DeepFork is replaced with these classifiers on our dataset to perform binary classification.
- **K Nearest Neighbor:** K - Nearest Neighbor algorithm assigns class membership to each input sample using majority voting. The sample is assigned to the class which is most common among its K nearest neighbors(Keller et al. 1985). In our case, the distance measure is the euclidean distance between samples, where each sample is a feature vector of d-dimension.
- **Linear SVM:** In an SVM model input samples are represented as points in space and the model tries to detect a hyperplane which can segregate categories of inputs by a clear gap that is as wide as possible (Cortes, and Vapnik 1995).
- **Decision Tree**: By recursively partitioning input samples into multiple classes based on some criteria, a decision tree is built. The criteria used for partitioning are based on the quality of the partition attained(Kohavi and Ron 1996). In our problem, the decision tree is build based on the values of individual features from the d-dimensional feature vectors in the training set.
- **Naive Bayes:** Bayesian classifiers assign the most likely class to a given sample described by its feature vector. The naive Bayes classifier greatly simplifies learning by assuming that features are independent of given class, which can be formulated as $P(X|C) = \prod_{i=1}^{n} P(X_i|C)$, where $X = (X_1, X_2, ......, X_n)$ (Rish et al. 2001].
- **Random Forest**: Random forests are combination of tree predictors such that each tree depends on the values of a random vector sampled independently and with the same distribution for all trees in the forest(Breiman, and Leo 2001).
- **Extremely Randomized Tree**: A new tree-based ensemble method for supervised classification and regression problems. It essentially consists of randomizing strongly both attribute and cut-point choice while splitting a tree node. In the extreme case, it builds totally randomized trees whose structures are independent of the output values of the learning sample(Geurts et al. 2006).
- **RBF SVM**: In addition to regular SVM Kernel, parameters Γ and C are used to know the how far the influence of a single training example reaches where highest value being too close and least values being too far, and to know misclassification of training examples(Chang et al. 2010).

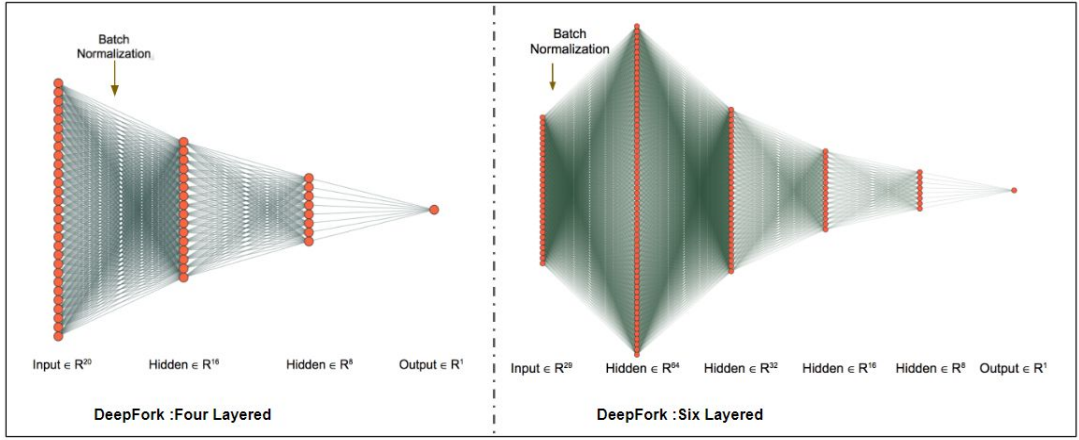

Figure 5: DeepFork Neural Network Architecture with 4 and 6 layers.

## 4 Empirical Study

In this section, we discuss the experimental setup and performance measurements for DeepFork. Later, we detailed about the results achieved.

### 4.1 Experimental Setup

We experimented with two variants of the deep feedforward neural network model with 4 and 6 fully connected layers. In both the variants, initial layer is a batch normalization layer(*only technique employed to normalize the input data*), followed by fully connected layers as shown in Figure 5. We experimented with Adam(Kingma et al. 2014) and Stochastic Gradient Descent (Bottou and Leon 2010) optimizers with various batch size(*32-256*), epochs (*50 - 250*), and learning rate(*1e-2, 1e-5*) settings and the best setup is mentioned in Table 1.

Table 1 : Best Parameter Setup for 4 and 6 Layered Fully Connected Deep Feed-Forward Neural Networks

| **Model Type** | **Epochs** | **Batch Size** | **Optimizer** | **Learning Rate** |
| --- | --- | --- | --- | --- |
| 4 Layered - Fully Connected NN Model | 100 | 64 | Adam | 1e-3 |
| 6 Layered - Fully Connected NN Model | 250 | 128 | Adam | 1e-4 |

Python based keras deep-learning framework, (Chollet and Francois 2017), is used for developing our models. The computational support used is an Amazon elastic compute cloud instance of type t2.medium: has 2 CPUs and 4GB Memory in-built (Jackson et al. 2010).

### 4.2 Performance Measurement

To evaluate the performance of our approach, initially, we compared the performance of our model using node specific features against traditional node similarity measures(*topological*) used for link prediction. Later, we employed different classification models on our dataset and then compared their performance with DeepFork. For the first set of experiments, we built a network using the nodes and edges in the training data as described above in Section 3.2. From this network, we extracted node similarity measures(*topological*) features which are 12-dimensional feature vectors and node-specific features which are 17-dimensional feature vectors to train the neural network model as described in Section 3.3. We compared the performance of our model under three settings:(1)Using node features (2)Using topological features (3)Using both topological and node features. We expect our model to perform better with node features when compared to topological features as we believe that node features incorporate useful information for link prediction. We also performed an ablation study under the first setting where we removed watch event information from the node features and reducing the feature size to **15**. This ablation study is performed to answer our research questions, if knowing that the user is watching a repository helps in predicting whether or not the user will fork the repository. The watch event is only a node level feature, therefore, it is not applicable when topological information is considered. For the second set of experiments, we compared our neural network model against other machine-learning models: K-Nearest Neighbor, Support Vector Machines, Decision Tree, Naive Bayes, Random Forest, and Extremely Random Tree. For all models, we used node specific features for training. Then, we computed the mean of the standard metrics:Accuracy, Precision, Recall and F1-Score, over 30 runs for every model.

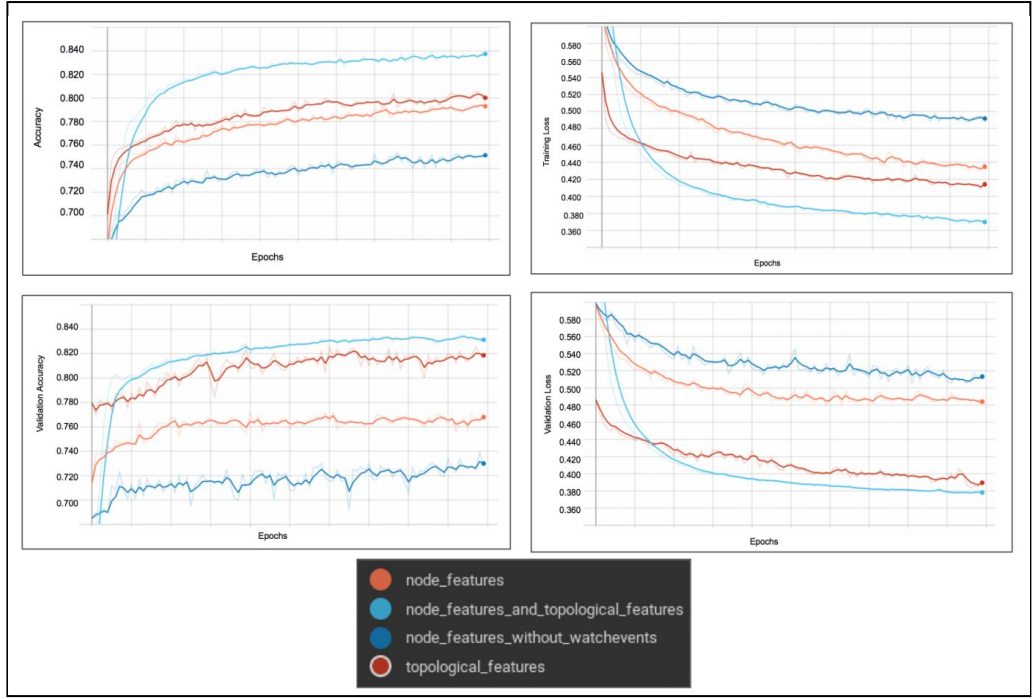

Figure 6. Plots forecasting evolution of training and validation loss and accuracy.

### 4.3 Results

The performance of training and validation for all DeepFork models are shown in Figure 6. These plots show how the neural networks learn over time by decreasing the training loss while increasing the classification accuracy. The statistical significance of the classifiers is measured using receiver optimizing curves(Davis et al. 2006), is shown in Figure 7; while Table 2 is the metric results from an individual experiment. From these results, we can infer that the standard machine learning classification models performed marginally better than chance. While random forest and decision tree models performed moderately, our DeepFork models performed best overall on this binary classification task. While training these models, we observed that Adam optimizer performed well over the stochastic gradient descent. Primary observation from these results is that the node level features outperform the topological features as expected. This shows that the topological features are more informative and are more useful for link prediction. We achieved the best results when we used both the node features and topological features and with these results, we can claim that our approach is very effective and we are able to predict if there is an information diffusion from a user to his *follower* with an accuracy of 75%. The results of the ablation study show that ignoring watch events from the input features drastically reduces the performance of the classifier, indeed signifying the importance of the "**Watch+Fork**" over just "**Fork**" in GitHub.

Table 2. Binary Classification Task Comparison among Benchmark Classifiers

|  |  | Precision | Recall | F1-Score |
| --- | --- | --- | --- | --- |

| Type of Classifier | Accuracy | +ve | -ve | +ve | -ve | +ve | -ve |
|---|---|---|---|---|---|---|---|
| Random | 0.502 | 0.494 | 0.509 | 0.502 | 0.497 | 0.496 | 0.501 |
| Naive Bayes | 0.517 | 0.514 | 0.520 | 0.626 | 0.407 | 0.565 | 0.456 |
| K Nearest Neighbor | 0.539 | 0.545 | 0.537 | 0.430 | 0.647 | 0.478 | 0.585 |
| Linear SVM | 0.519 | 0.165 | 0.333 | 0.333 | 0.666 | 0.220 | 0.444 |
| RBF SVM | 0.527 | **0.900** | 0.576 | 0.017 | **1.0** | 0.336 | 0.673 |
| Random Forest | 0.692 | 0.734 | 0.667 | 0.562 | 0.812 | 0.637 | 0.733 |
| Decision Tree | 0.691 | 0.701 | 0.682 | 0.670 | 0.711 | 0.685 | 0.696 |
| Extremely Randomized Tree | 0.675 | 0.689 | 0.664 | 0.639 | 0.711 | 0.663 | 0.687 |
| DeepFork - Node | 0.726 | 0.734 | 0.719 | 0.691 | 0.759 | 0.712 | 0.738 |
| DeepFork - Node - No Watch | 0.611 | 0.612 | 0.610 | 0.563 | 0.656 | 0.586 | 0.632 |
| DeepFork - Topological | 0.677 | 0.691 | 0.666 | 0.617 | 0.735 | 0.652 | 0.699 |
| DeepFork - Node -Topological - Watch | **0.752** | 0.757 | **0.747** | **0.742** | 0.762 | **0.750** | **0.754** |

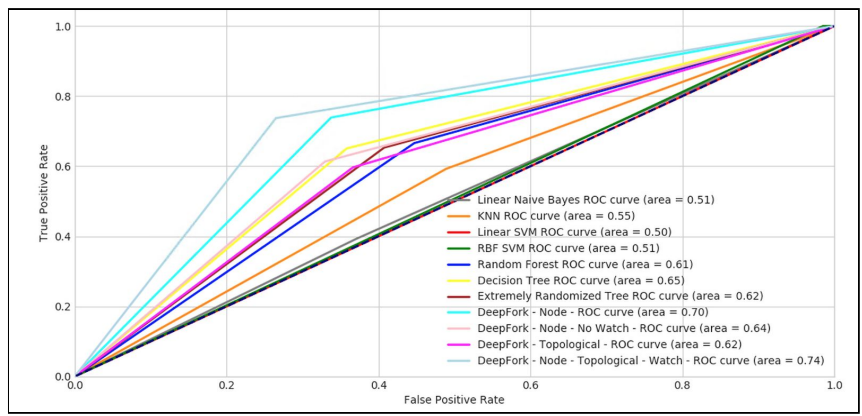

Figure 7. Plots representing Receiver Optimizing Curve to show performance of various classifiers

## 5 Conclusion

To predict the diffusion of information in complex social coding networks: GitHub, a deep feedforward neural network model:*DeepFork* is developed, that in turn aids us to understand the effects of human behavior.

- **DeepFork** measured **Influence:** *followees* play a big role while directing their *followers* to new repositories and providing them various opportunities such as to contribute, to improve knowledge, to build their affiliations and so on.
- **DeepFork** measured **Information Spread:** any thriving repository that has code or any piece of information is rapidly escalated across the network with the help of Fork and Watch events, as Fork depicts reputation/popularity and Watch depicts enthusiasm/interest.
- **DeepFork** measured **Evolution:** Modifications from the forked repository can either be contributed back to the original repository or give rise to new repository that leads to other software development. In our empirical studies, we observed that information diffusion can be detected by link prediction using supervised learning.

DeepFork has outperformed other machine learning models as they better learn the discriminative patterns from the input features. Furthermore, we are keen to explore repository patterns, which repositories serve as gatekeepers and why.

## Acknowledgement

*Funded by DARPA for Computational Simulation of Online Social Behavior (SocialSim); Solicitation Number: HR001117S0018; Office: Defense Advanced Research Projects Agency*

## Biographies:

**Ramya Akula** is a Ph.D student at UCF Complex Adaptive Systems Laboratory in Industrial Engineering and Management Systems at the University of Central Florida, Orlando, USA. She earned her bachelor's degree in Computer Science Engineering from Jawaharlal Nehru Technological University, India and masters degree in Computer science from Technical University of Kaiserslautern, Germany. She completed many software enhancement projects in industry and academia. Her research interests are intelligent systems, machine-human behaviorial computation and social network analysis.

**Niloofar Yousefi** is a Research Associate at UCF Complex Adaptive Systems Laboratory in Industrial Engineering and Management Systems at the University of Central Florida, Orlando, USA. Her research is focused on machine learning, computational learning theory, generalization bounds, kernel-based methods and Multi-Task Learning (MTL). Her research contributions are into theoretical aspect of MTL problem that can strongly interplay with experimental studies while potentially can push the boundaries of this learning setting and lead to designing more accurate models compared to both not only traditional single task learning approaches, but also existing MTL models.

**Ivan Garibay** is an Assistant Professor and Director of UCF Complex Adaptive Systems Laboratory in Industrial Engineering and Management Systems at the University of Central Florida, Orlando, USA. He has been leading diverse technology, innovation funded research projects as the Chief Information Officer for UCF research division for over 15 years. He is also the founding director of two very successful initiatives: Master of Science in Data Analytics(MS program) and I-Corps, a UCF-Site program for entrepreneurial education. His current active research is on modeling of online social behavior to study influence, misinformation and propaganda in social networks.